\documentclass[
	twocolumn,
	twocolappendix,
	]{aastex631}

\let\oldtablenum\tablenum 
\let\tablenum\relax
\usepackage{siunitx}
\let\tablenum\oldtablenum

\usepackage{amsmath}
\usepackage{mathtools} 
\usepackage{booktabs} 
\usepackage{bm} 

\usepackage{dcolumn} 
\newcommand\mcL[1]{\multicolumn{1}{c}{#1}} 

\allowdisplaybreaks 

\newcommand{\Textcite}{\Citet}
\newcommand{\parencite}{\citep}

\usepackage[inline]{enumitem} 
\newenvironment{enuminline}{\begin{enumerate*}[label=(\roman*)]}{\end{enumerate*}} 

\renewcommand{\dh}{\partial}
\renewcommand{\d}{\mathrm{d}}
\newcommand{\D}{\mathrm{D}} 

\newcommand{\dive}{\bm{\nabla}\!\cdot}
\newcommand{\curl}{\bm{\nabla}\!\times\!}
\newcommand{\grad}{\bm{\nabla}}

\newcommand{\defn}{\equiv}
\newcommand{\abs}[1]{\left|#1\right|}
\newcommand{\op}[1]{\operatorname{#1}}

\newcommand{\scalesAs}{\sim}

\renewcommand{\vec}[1]{\boldsymbol{#1}}
\newcommand{\cross}{\times}
\newcommand{\uvec}[1]{\boldsymbol{\hat{#1}}} 
\newcommand{\R}{\ensuremath{\operatorname{Re}}} 
\newcommand{\Nu}{\ensuremath{\operatorname{Nu}}} 
\newcommand{\Ra}{\ensuremath{\operatorname{Ra}}} 
\newcommand{\Ma}{\ensuremath{\operatorname{Ma}}} 
\newcommand{\RaF}{\ensuremath{\operatorname{Ra}_\mathrm{F}}} 
\newcommand{\RaSGS}{\ensuremath{\operatorname{Ra}_\text{SGS}}} 
\newcommand{\pencil}{Pencil} 

\newcommand{\dyad}[1]{\overset{\text{\tiny$\bm\leftrightarrow$}}{#1}}
\newcommand{\cool}{\text{cool}} 
\newcommand{\sclk}{{\widetilde{k}}}
\newcommand{\sclom}{{\widetilde{\omega}}}

\newcommand{\FT}[1]{\widetilde{#1}} 
\newcommand{\enth}{\text{enth}} 
\newcommand{\chiSGS}{\chi_{\rm SGS}}
\newcommand{\meanBrxy}[1]{\left<#1\right>_{xy}}
\newcommand{\meanBrxyt}[1]{\left<#1\right>_{xyt}}
\newcommand{\Fbot}{F_\text{bot}}

\shorttitle{Strengthening of the f mode due to magnetic fields}
\shortauthors{Kishore, Singh,  K\"apyl\"a, \& Roth}

\begin{document}

\title{Strengthening of the f mode due to subsurface magnetic fields in simulations of convection}
\correspondingauthor{G. Kishore}

\author[0000-0003-2620-790X]{G. Kishore}
\affiliation{IUCAA, Post Bag 4, Ganeshkhind, Pune 411007, India}
\email{kishoreg@iucaa.in}

\author[0000-0001-6097-688X]{Nishant K Singh}
\affiliation{IUCAA, Post Bag 4, Ganeshkhind, Pune 411007, India}
\email{nishant@iucaa.in}

\author[0000-0001-9619-0053]{Petri K\"apyl\"a}
\affiliation{Institut f\"ur Sonnenphysik (KIS), Georges-K\"ohler-Allee 401a, 79110 Freiburg im Breisgau, Germany}
\email{pkapyla@leibniz-kis.de}

\author{Markus Roth}
\affiliation{Th\"uringer Landessternwarte Tautenburg and Friedrich-Schiller-Universit\"at Jena, Germany}
\email{mroth@tls-tautenburg.de}

\begin{abstract}
	Previous studies have found that localized strengthening of the f mode precedes the emergence of active regions on the Sun by one day to three days.
	To help interpret these observations, we have performed nonlinear simulations of convection with imposed magnetic fields at different depths.
	We find that the f mode is strengthened when a super-equipartition magnetic field is imposed near the top of the domain.
	However, neither a magnetic field of equal strength near the bottom of the domain nor an equipartition magnetic field near the top of the domain have a significant effect.
	Our results suggest that the magnetic precursors of active regions are present near the surface of the Sun for much longer than would be expected if active regions were formed by flux tubes rising from deep within the convection zone.
	Application to observations should account for the fact that the effects we observe are transient.
\end{abstract}


\section{Introduction}
The Sun supports a wide variety of waves.
Depending on the nature of the restoring force, the corresponding modes \citep{LeiSte81} are classified into various branches.
The eigenfunctions of each type of mode have characteristic dependences on depth, and are thus sensitive to different layers of the Sun.
Helioseismology uses observations of the amplitudes and frequencies of these modes at the solar surface to infer the properties of deeper layers \parencite{Chr02, Tho06}.

At a free surface, the incompressible deep-water equations support horizontally travelling waves whose amplitudes decay exponentially with depth over a scale proportional to their wavelength \parencite[e.g.][sec.~3.2 and exercise 3.2]{Ach90}.
These waves are also supported at interfaces with a finite density contrast.
In the stellar context, the associated modes are usually referred to as f (fundamental) modes.
Since their eigenfunctions decay rapidly with depth (at least for large horizontal wavenumbers), these modes are only expected to be affected by the near-surface layers of the star.

\Textcite{SinRaiBra16} and \textcite{WaiRotSin23} have found that localized strengthening of the solar f mode precedes the emergence of an active region by one day to three days.\footnote{
\Textcite{KorKorOls22}, on the other hand, find that there is no significant effect.
Appendix \ref{section: continuum power law slope} discusses a possible reason for this disagreement.
}
For the large horizontal wavenumbers considered in these observations, the eigenfunction of the f mode is expected to decay with depth over a scale of a few \si{Mm}.
For example, \citet{SinRaiBra16} considered f modes with $k R_\sun > 1200$, which decay over less than \SI{0.6}{Mm}.
This suggests that the precursors of active regions can be present in the upper few \si{Mm} of the convection zone (CZ) up to three days before their emergence.

Sunspot formation is often explained by appealing to magnetic flux tubes buried at the bottom of the CZ, approximately \SI{200}{Mm} below the surface \citep[for a review, see][]{charbonneau2020}.
Such flux tubes may rise either due to magnetic buoyancy \citep{Par75}, or due to kink instabilities \citep[p.~577]{Bab61}.
Rising flux tubes are expected to cross the top \SI{10}{Mm} of the CZ (around 11 pressure scale heights) in a few hours \parencite{CheRemTit10}.
This is difficult to reconcile with the aforementioned observations.

Going beyond the qualitative reasoning described above, one would like to use mode amplitudes to quantitatively constrain the locations, strengths, and configurations of subsurface magnetic fields.
However, analytical predictions of the effects of magnetic fields on helioseismic modes typically only consider simple magnetic field configurations, and are limited to studying frequency shifts \parencite[e.g.][]{MilAllRob92, TriMit22, RuiFul23}.
The general formalism presented by \textcite{KieSchRot17} and \textcite{KieRot18} also only deals with frequency shifts.

\Textcite{SinBraRhe14, SinBraChi15, SinRaiKap20} performed 2D simulations where turbulence is driven by a prescribed forcing function in a piecewise isothermal background.
Imposing magnetic fields of different configurations, they observed changes in the frequencies and the amplitudes of various modes.
\Textcite[sec.~4.2]{SinBraChi15} found that a strong, uniform, horizontal magnetic field suppresses the amplitude of the f mode at all wavenumbers.
In their setup, they found that the mode amplitude is correlated with the RMS velocity of the flow; it is only through the latter that the magnetic field affects the mode amplitudes.
On the other hand, \textcite[sec.~3.3]{SinRaiKap20} found that when the magnetic field is localized, the f mode is strengthened at large wavenumbers, but only when the magnetic field is near the surface.

The amplitude of a particular mode depends on the spectral properties of the forcing function used (\citealp[sec.~4.1]{batchelor1953homoturb}; \citealp[sec.~V.F]{Chr02}).
In the case of the Sun, the modes are excited by convective flows, which may themselves be affected by the presence of a magnetic field.
Thus, while the analytical studies and idealized forced-turbulence simulations described above can be used to understand how the frequencies of various modes are affected by magnetic fields, it is unclear if their findings on the amplitudes of various modes can be applied to convective systems.
The amplitudes are sensitive to the nature of the fluid flows, and so one has to perform nonlinear simulations of stratified convection to capture the relevant effects.
While many authors have observed modes (f, p, and Rossby) in simulations of convection \parencite[e.g.][]{ZhaGeoKos07, BekCamGiz22_2, WaiZhaKit23, BluHinMat24}, we are not aware of any studies of magnetic effects on modes in such simulations.

In this study, we use simulations in which convection is driven by a prescribed cooling function at the top of the domain.
Peaks of the horizontally and temporally Fourier-transformed vertical velocity from such simulations can be associated with f, p, and g modes.
Imposing horizontal magnetic fields localized at different depths, we probe their effects on the amplitude of the f mode near the top of the domain.

In section \ref{section: numerical}, we specify our simulation setup.
Section \ref{section: definitions} defines quantities in terms of which we report our results.
Section \ref{section: non-magnetic} describes the statistically steady state of our non-magnetic simulation.
In section \ref{section: f mode str}, we report the effects of imposed magnetic fields on the f mode.
We summarize our work in section \ref{section: conclusions}.

\section{Numerical setup}
\label{section: numerical}
\subsection{Domain and evolution equations}
We consider a Cartesian domain, periodic in the $x$ and $y$ directions.
Gravity points along the $-\uvec{z}$ direction.
The domain extends vertically between $-0.45 \, l < z < 1.05 \, l$, where $l$ is the depth of the initially unstable region.

We solve the following evolution equations:
\begin{align}
	\begin{split}
		\frac{\D \ln\rho}{\d t} ={}& - \dive\vec{u}
	\end{split}
	\\
	\begin{split}
		\rho \, \frac{\D \vec{u}}{\d t}
		={}&
		- \grad p
		- \rho g \uvec{z}
		+ \vec{J} \cross \vec{B}
		+ \dive{\!\left( 2 \rho \nu \dyad{S} \right)}
	\end{split}
	\\
	\begin{split}
		\rho T \, \frac{\D s}{\d t}
		={}&
		q
		- \dive{\!\left( \vec{F}_\text{rad} + \vec{F}_\text{SGS} \right)}
		+ 2 \rho\nu S_{ij} S_{ij}
	\end{split}
\end{align}
where
\begin{align}
	\begin{split}
		\vec{B} \defn{}& \curl\vec{A}
	\end{split}
	\\
	\begin{split}
		\vec{J} \defn{}& \frac{ \curl\vec{B} }{\mu_0}
	\end{split}
	\\
	\begin{split}
		S_{ij}
		\defn{}&
		\frac{1}{2} \left( \dh_iu_j + \dh_ju_i - \frac{2}{3} \delta_{ij}\dive\vec{u}\right)
	\end{split}
\end{align}
with
$\rho$ being the density;
$\vec{u}$ the velocity;
$\D/\d t \defn \dh/\dh t + \vec{u} \cdot \grad$ the convective derivative;
$p$ the pressure;
$g$ the acceleration due to gravity (with $g > 0$);
$\nu$ the kinematic viscosity;
$s$ the specific entropy;
$\vec{A}$ the magnetic vector potential; and
$\mu_0$ the magnetic permeability.
A similar setup (without magnetic fields) has been used by \citet{Kap19_2, kapyla2021Pr}.

We choose units such that $g = C_P = l = \rho_0 = 1$, where
$C_P$ is the specific heat at constant pressure,
and
$\rho_0$ is the initial density at the top of the domain.
With these choices, the unit for length is $l$;
mass is $\rho_0 l^3$;
time is $\sqrt{l/g}$;
temperature is $gl/C_P$; and
magnetic field is $\sqrt{\mu_0 \rho_0 g l}$.
We assume the kinematic viscosity is a constant ($\nu = 10^{-4} \, l^{3/2} g^{1/2}$), and neglect the bulk viscosity (omitted from the expressions above).

In the evolution equation for the entropy, $\vec{F}_\text{rad}$ is the radiative flux determined by the Kramers opacity formula (i.e.\@ the thermal conductivity is itself a function of the temperature and the density), such that
\begin{equation}
	\vec{F}_\text{rad} = - K_0 \, \rho^{-2} \, T^{13/2} \, \grad T
	\label{eq: F_rad}
\end{equation}
where $T$ is the temperature.
\Textcite{kapyla2017simulation} have studied how such a prescription differs from more common choices.
We set $K_0 \approx 1.41 \, \rho_0^3 C_P^{15/2} l^{-5} g^{-6} $. 

Following \textcite{kapyla2021Pr}, for numerical reasons, we
also add a subgrid-scale (SGS) diffusivity that acts on the fluctuations of the entropy about its horizontal average.
The SGS diffusivity is set to $\chiSGS = 10^{-4} \, l^{3/2} g^{1/2}$, and the entropy flux due to it is given by
\begin{equation}
		\vec{F}_\text{SGS}
		\defn
		- \rho T \chiSGS \grad s'
	\,,\quad
		s' \defn s - \meanBrxy{s}
	\,.
	\label{eq: F_SGS}
\end{equation}
A cooling function, $q$, of the form
\begin{equation}
	q = - \frac{ \rho \left( c_s^2 - c_\cool^2 \right) }{\tau_\cool } \op{\Theta}{\!\left( \frac{ z - z_\cool }{w_\cool } \right)}
	\label{eq: cooling function}
\end{equation}
with $\op{\Theta}(z) \defn \left( 1 + \op{erf}(z) \right)/2$, $z_\cool = l$, $c_\cool^2 = 0.09 \, gl$, $w_\cool = 0.025 \, l$, and $\tau_\cool = \left( 1/27 \right) \sqrt{l/g}$ is also added to the entropy equation.
This forces the adiabatic sound speed ($c_s$) in $z > z_\cool$ to $c_\cool$ over a timescale $\tau_\cool$.

The fluid is chosen to be an ideal gas, whose equation of state is
\begin{equation}
	p = \rho R T
	= \rho_0 R T_0 \left( \frac{\rho}{\rho_0} \right)^{\gamma} e^{\gamma s/C_P}
\end{equation}
where $R \defn C_P - C_V$ and $\gamma \defn C_P/C_V$, with $C_V$ being the specific heat at constant volume.
We set $\gamma = 5/3$, corresponding to a monoatomic gas.
Note that the zero point of the entropy is determined by the constant $T_0$, which we choose such that $\left( \gamma - 1 \right) C_P T_0 = c_\cool^2$.

We prescribe a time-independent magnetic field by choosing the following form for $\vec{A}$:
\begin{equation}
	A_y =
	\begin{dcases}
		\frac{2 C}{\pi} \left( z_2 - z_1 \right) & z < z_1 \\
		\frac{C}{\pi} \left( z_2 - z_1 \right) \left( 1 + \cos{\!\left[ \pi \, \frac{z - z_1}{z_2 - z_1} \right]} \right) & z_1 \le z < z_2 \\
		0 & z_2 \le z
	\end{dcases}
\end{equation}
with $A_x = A_z = 0$.
This is constructed in such a way that the resultant magnetic field is in the $\uvec{x}$ direction, is a function of only $z$, and is localized in the region $z_1 < z < z_2$.
$C$ is the maximal value of the magnetic field.
Table \ref{table: params and diagnostics} gives the values of $C$, $z_1$, and $z_2$ for all the simulations we consider,
while the resulting profiles of $B_x$ are plotted in figure \ref{helio.conv-mag.SBT: fig: bx ST}.

\begin{table*}
	\centering
	\begin{tabular}{
		c
		D{.}{.}{1.1}
		D{.}{.}{1.1}
		D{.}{.}{1.2}
		r@{}c@{}l
		c
		D{.}{.}{1.1}
		c
		D{.}{.}{1.1}
		c
		D{.}{.}{1.2}
		D{.}{.}{1.1}
		l
		}
		\toprule
			Name   &  \mcL{$C$}                          & \mcL{$z_1$} & \mcL{$z_2$}   & \multicolumn{3}{c}{Time range}      & Realizations & \mcL{$\RaSGS$}               &   \mcL{$\R$} &   \mcL{$\Ra$}                   &   \mcL{$\Nu$} &   \mcL{$\Ma_\text{max}$} & \mcL{$\RaF$}    & Comment           \\
			{}     &  \mcL{[$\sqrt{\mu_0 \rho_0 g l}$]}  & \mcL{[$l$]} & \mcL{[$l$]}   & \multicolumn{3}{c}{[$\sqrt{l/g}$]}  & {}           & \multicolumn{1}{c}{[$10^6$]} &     {}       &   \multicolumn{1}{c}{[$10^7$]}  &   {}          &   {}                     & \mcL{[$10^8$]}  &                   \\
		\midrule
			NF     &                                0    & {}          & {}            & 300&--&1300                         &    4         &              1.6             &          265 &                      7.5        &            15 &                     0.12 &            4.6  & No Field          \\
			ST     &                                0.5  & 0.4         & 1.05          & 0&--&250                            &    8         &              2.7             &          373 &                     13          &            13 &                     0.16 &            5.2  & Strong Top        \\
			ST     &                                0.5  & 0.4         & 1.05          & 250&--&500                          &    8         &              1.9             &          336 &                      9.5        &            14 &                     0.15 &            5.2  & Strong Top        \\
			SB     &                                0.5  & -0.45       & 0.2           & 0&--&250                            &    6         &              1.6             &          268 &                      7.5        &            15 &                     0.12 &            4.7  & Strong Bottom     \\
			ET     &                                0.14 & 0.4         & 1.05          & 0&--&250                            &    6         &              1.6             &          273 &                      7.6        &            14 &                     0.13 &            4.6  & Equipartition Top \\
		\bottomrule
	\end{tabular}
	\caption{
		Summary of the discussed simulations, along with diagnostics computed from a single realization of each configuration.
		For the magnetic cases, $t=0$ corresponds to the time when the magnetic field is imposed.
		The diagnostics have been computed by averaging over the entire interval indicated.
		The same interval was used to compute the power spectra.
		Recall that for all the runs, $\nu = \chiSGS$.
		}
	\label{table: params and diagnostics}
\end{table*}

These evolution equations are solved using the \pencil{} code \citep{Pencil2021}.\footnote{
\url{https://pencil-code.nordita.org}
}
Spatial derivatives are discretized using a sixth-order finite difference scheme, while the equations are evolved in time using a third-order Runge-Kutta method.
Upwinding is used for advection of the density, the velocity, and the entropy.
We use $N_x\times N_y \times N_z = 1152^2\times 288$ grid points with $L_x = L_y = 16 l$.
In appendix \ref{appendix: grid-independence}, we show that the convective flows are well-resolved with the chosen grid spacing.

\subsection{Boundary conditions}
The vertical boundaries are made free-slip, impenetrable, and perfect electric conductors.
The total energy flux at the bottom is constrained to be $\Fbot \defn 5 \times 10^{-4} \, \rho_0 g^{3/2} l^{3/2}$, while the temperature at the top is constrained to be $T_0$.

\subsection{Initial conditions}
\label{section: initial conditions}

To have decent resolution in $k$-space, we require the aspect ratio (horizontal/vertical dimension) of the domain to be large.
However, running such simulations till they reach a statistically steady state is expensive.
While an accelerated evolution scheme has been proposed to reduce the amount of time required to reach steady state \citep{AndBroOis18}, implementing it in an existing code seems complicated.
We use a different approach, described below.

We first run a non-magnetic simulation with $L_x = L_y = l$ for $10^{4} \sqrt{l/g}$ time units.
For this simulation, the initial condition is piecewise polytropic, with the polytropic index being given by
\begin{equation}
	\begin{dcases}
		3.25 & z < 0
		\,,\\
		1.5 & 0 < z < l
		\,,\\
		\infty & l < z
		\,.
	\end{dcases}
\end{equation}
The temperature at the top is set to $T_0$.
The velocity at each grid point is randomly chosen such that each of its components are uniformly distributed between $\pm 10^{-4} \sqrt{gl}$.

A snapshot (an array containing all the state variables at each grid point) from this simulation is then periodically tiled to generate the initial condition for the non-magnetic simulation.
To break the symmetry of the initial condition, low-amplitude grid-scale noise (with each of its components at each grid point uniformly distributed between $\pm 10^{-3} \sqrt{gl}$) is added to the velocity field.
This leads to the symmetry of the initial condition being broken within a few turnover times.
We produce four realizations of the non-magnetic case, differing only in the initial random component of the velocity field added during tiling.
For all our analyses, we ignore data from the first $300 \, \sqrt{l/g}$ time units of the larger non-magnetic simulations.

The magnetic simulations are initialized from snapshots  of the non-magnetic simulations.
To ensure statistical independence of the different realizations of a particular magnetic configuration, we ensure that even when two such realizations are initialized from the same non-magnetic realization, the snapshots have a minimum time-separation of $1000 \, \sqrt{l/g}$.

\subsection{Simulation output}
The horizontal Fourier transform of the vertical component of the velocity is saved at all depths for $\abs{k_x},\abs{k_y} \le \gamma g/c_\text{top}^2$ at intervals of $0.1 \sqrt{l/g}$ time units ($0.5 \sqrt{l/g}$ for the non-magnetic case).
The temporal Fourier transform is later performed during post-processing.

\subsection{Sources of error}
\label{section: error estimation}

We expect the two main sources of error to be
\begin{enuminline}
	\item the finiteness of the integration time; and
	\item the spatial discretization used.
\end{enuminline}
We estimate the error due to the former by considering multiple realizations of the same configuration.
Appendix \ref{appendix: grid-independence} studies how the spatial discretization affects the results of this study.

Unless otherwise mentioned, realization-dependent quantities are estimated by averaging over all realizations; we consider the error in this estimate to be $\sigma/\sqrt{N}$, where $\sigma$ is the standard deviation between different realizations and $N$ is the number of realizations.
Note that an estimate of this kind does not account for the error due to the spatial discretization.

\section{Definitions, conventions, and models}
\label{section: definitions}
\subsection{Dimensional scales}

As the length and frequency scales, we take
\begin{equation}
		L_0
		\defn
		\frac{ c_\text{top}^2 }{ \gamma g }
		\,,
		\quad
		\omega_0
		\defn
		\frac{ g }{ c_\text{top} }
		\,.
\end{equation}
Note that $L_0$ would be the pressure scale height at the top of the domain if it were in hydrostatic equilibrium; $\omega_0$ is proportional to the acoustic cutoff frequency in an isothermal background.

Using these, we define dimensionless wavenumbers and frequencies as
\begin{equation}
		\sclom
		\defn
		\frac{\omega}{\omega_0}
		\,,
		\quad
		\sclk_i
		\defn
		k_i L_0
		\,.
\end{equation}

\subsection{Power spectrum}
Defining $\FT{u}(\omega, k_x, k_y, z)$ as the discrete Fourier transform\footnote{
For a periodic function, $f(x)$, defined on $-L/2 \le x < L/2$, we use the convention that its Fourier transform, $f_n$, is given by
$\frac{1}{L} \int_{-L/2}^{L/2} f(x) \, e^{-2\pi i n x/L} \, \d x$.
} of $u_z(t,x,y,z)$ in time and the two horizontal directions, we define the power as
\begin{equation}
	P(\sclom, \sclk_x, \sclk_y, z) \defn \abs{\FT{u}(\omega, k_x, k_y, z)}^2
	\,.
	\label{helio.conv-mag.defn: P}
\end{equation}

Previous studies (\citealp[eq.~5]{SinBraRhe14}; \citealp[eq.~12]{SinBraChi15}; \citealp[eq.~8]{SinRaiKap20}) used the absolute value of the Fourier-transformed velocity, rather than its square.
Our choice has been used in observational studies \parencite[e.g.][p.~2]{SinRaiBra16}.
Note that it is the squared magnitude of the Fourier transform that is expected to show a Lorentzian mode profile (\citealp[sec.~4.1]{batchelor1953homoturb}; \citealp[sec.~V.F]{Chr02}).

\subsection{Fitting modes}
\label{maghelio: section: mode model}

Consider the power spectrum, defined in equation \ref{helio.conv-mag.defn: P}, at fixed $\sclk_x$, $\sclk_y$, and $z$.
We model it as the sum of a power-law (the continuum) and a number of Lorentzians (each corresponding to a mode).
For nonlinear least-squares fitting, we use an implementation of a trust-region reflective algorithm, provided by SciPy \citep{scipy2020}.
The errors in $P(\sclom)$ (estimated as described in section \ref{section: error estimation}) are propagated in the usual way to estimate errors in the fit parameters.
Appendix \ref{appendix: fitting modes} gives more details on these steps.

\subsection{Mode mass}
\label{section: mode mass defn}
Given a particular mode, if the resolution in $\omega$ is not much smaller than the width of that mode, the peak value of $\FT{u}(\omega)$ is sensitive both to the resolution and to any additional smoothing kernels that one may employ.
As a measure of the strength of a particular mode, \textcite[eq.~19]{SinBraChi15} defined a `mode mass', which is proportional to $\int \Delta \FT{u} \, \d \omega$, where $\Delta\FT{u}$ denotes $\FT{u}$ after removal of a `continuum' (and possibly other modes of disinterest).

We define the mode mass as
\begin{equation}
	\mu \defn \sum_\sclom \Delta P(\sclom)
\end{equation}
where $\Delta P(\sclom)$ is the component corresponding to the mode of interest in the fit to the power spectrum (section \ref{maghelio: section: mode model}).
By virtue of Parseval's theorem \citep[problem 10.24]{Bracewell3rd}, $\mu$ defined in this way measures the temporally averaged specific kinetic energy associated with a particular mode; it is independent of the integration time if the system is in a statistically steady state.
The analogous quantities defined by \citet[eq.~19]{SinBraChi15} and \citet[eq.~9]{SinRaiKap20} do not have this property.\footnote{
\Textcite[p.~207]{SinRaiKap20} allude to this.
}\textsuperscript{,}\footnote{
This can be understood by noting that if $\sum_\omega P$ is independent of the integration time (say $T$), one can write $P \scalesAs 1/T \implies \abs{\FT{u}_z} \scalesAs 1/\sqrt{T} \implies \sum_\omega \abs{\FT{u}_z} \scalesAs \sqrt{T}$.
}

The errors in the fit parameters (section \ref{maghelio: section: mode model}) are used to estimate the error in the mode mass by computing the mode mass for $10^4$ realizations of the fit parameters (assumed to be normally distributed, with the distribution truncated at zero for parameters which are expected to be non-negative).
The upper and lower errors in the mode mass are then estimated from the 84.1 and 15.9 percentile values of the mode mass (these correspond to 1-$\sigma$ deviates for a normal distribution).

\subsection{Fluxes}
The averaged total energy density (kinetic + internal + gravitational) is transported by the following fluxes:
\begin{align}
		\vec{F}_\enth
		\defn{}&
		\meanBrxyt{\rho C_P \vec{u} T}
	\\
		\vec{F}_\text{kin}
		\defn{}&
		\meanBrxyt{ \rho u^2 \vec{u} / 2 }
	\\
		\vec{F}_\text{visc}
		\defn{}&
		- \meanBrxyt{ 2 \rho \nu \vec{u} \cdot \dyad{S} }
	\\
		\vec{F}_\cool
		\defn{}&
		- \uvec{z} \int_{z_\text{bot}}^z \meanBrxyt{ q } \d z
\end{align}
along with the radiative and SGS fluxes defined in equations \ref{eq: F_rad} and \ref{eq: F_SGS}.
Above, $\meanBrxyt{\Box}$ denotes the horizontal and temporal average of $\Box$ (in practice, the temporal average is taken over a few turnover times), while $q$ is given by equation \ref{eq: cooling function}.

\subsection{Diagnostics}

We define the Rayleigh number ($\Ra$), which describes the degree of supercriticality of convection, as
\begin{equation}
	\Ra
	\defn
	\frac{ g \abs{\Delta s} L^3 }{ C_P \nu \chi }
\end{equation}
where $L$ is the depth over which the entropy gradient remains negative (i.e.\@ excluding the stably stratified layers at the top and the bottom of the domain);
$\Delta s$ is the integral over this depth of the horizontally and temporally averaged vertical entropy gradient;
and
$\chi$ is the thermal conductivity (in our case, that arising from the Kramers opacity formula, such that $\chi = K_0 \rho^{-3} T^{13/2} C_P^{-1}$).
We choose the value of $\chi$ corresponding to the midpoint of the region where the entropy gradient is negative.
A similar quantity can be defined using the subgrid diffusivity:
\begin{equation}
	\RaSGS
	\defn
	\frac{ g \abs{\Delta s} L^3 }{ C_P \nu \chiSGS }
	\,.
\end{equation}
We define the Reynolds number, $\R$, which measures how turbulent the system is, as
\begin{equation}
	\R \defn \frac{ \op{max}_z[ u_\text{rms}(z) ] \, L }{\nu}
	\,.
\end{equation}
The contribution of convection to the total heat transport is characterized by the Nusselt number, $\Nu$, which is the ratio of the total energy flux to that carried by radiation.
We also report the peak value of the Mach number:
\begin{equation}
	\Ma_\text{max} \defn \sqrt{ \op{max}_z[ \meanBrxyt{u^2} /  \meanBrxyt{c_s^2} ]}
\end{equation}

The system can also be characterized by the flux-based Rayleigh Number:
\begin{equation}
	\RaF
	\defn
	\frac{g l^4 \Fbot}{C_P \rho T \nu \chi^2}
\end{equation}
where $\rho$, $T$, and $\chi$ are measured at $z = 0$.
Note that this is akin to the product of a Rayleigh number and the Nusselt number.

\section{The non-magnetic case}
\label{section: non-magnetic}

\begin{figure*}
	\centering
	\includegraphics{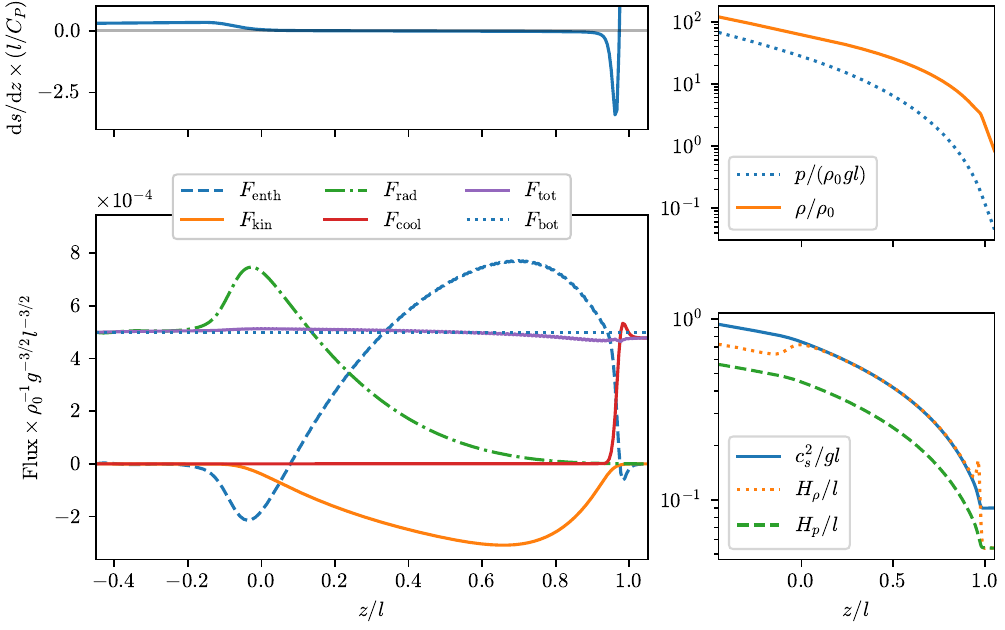}
	\caption{
	Horizontally and temporally averaged vertical fluxes and other quantities in a realization of the nonmagnetic (NF) simulation.
	The constant flux marked as $\Fbot$ is that used to set the boundary condition for the entropy at the bottom of the domain.
	$\vec{F}_\text{tot}$ is the sum of $\vec{F}_\enth$, $\vec{F}_\text{kin}$, $\vec{F}_\text{rad}$, and  $\vec{F}_\cool$; contributions due to $\vec{F}_\text{SGS}$ and $\vec{F}_\text{visc}$ have not been included.
	}
	\label{helio.conv-mag.hydro: fig: background + fluxes}
\end{figure*}

\begin{figure*}
	\centering
	\includegraphics{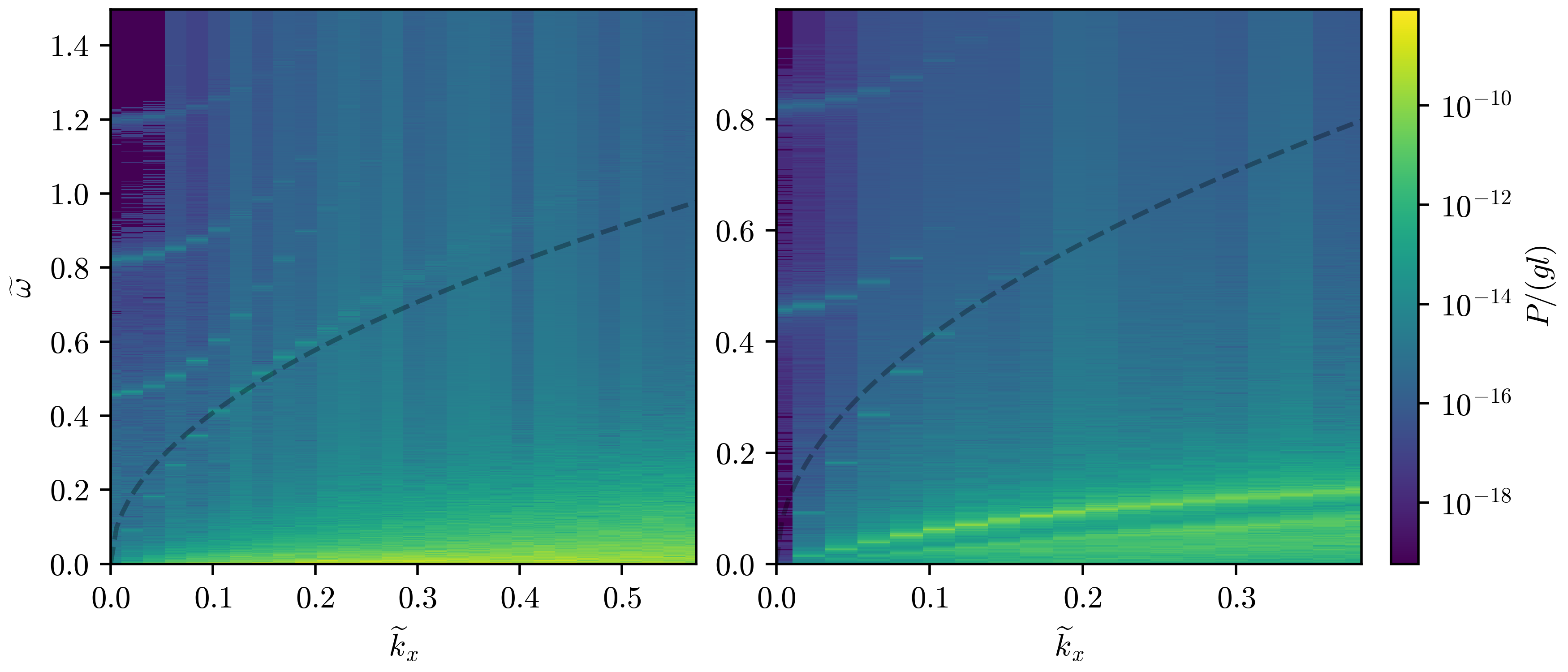}
	\caption{
	Diagnostic $k$-$\omega$ diagrams generated by averaging over  four realizations of the non-magnetic (NF) case for (left) $z=l$; and (right) $z=-0.1l$.
	In both cases, we have set $\sclk_y = 0$.
	The dashed black line shows the theoretical dispersion relation of the f mode.
	}
	\label{fig: komega NF}
\end{figure*}

Figure \ref{helio.conv-mag.hydro: fig: background + fluxes} shows the vertical components of various fluxes of the energy, along with the density, the squared sound speed, the vertical entropy gradient, and the RMS velocity (all horizontally and temporally averaged) as a function of depth
in a single realization of the NF case.
The temporal averages have been taken over six turnover times in the statistically steady phase of the simulation. 
Table \ref{table: params and diagnostics} lists various diagnostics that characterize the simulation.
The stratification becomes stable for $z < 0.36 l$, but still remains close to adiabatic for $0 \lesssim z < 0.36 l$.
Like the Kramers runs discussed by \textcite{kapyla2017simulation}, our setup contains an extended Deardorff zone, where the convective flux is positive (directed upward) despite the stratification being stable.
Below this is an overshoot zone, where the convective flux becomes negative.
The prescribed cooling function leads to almost isothermal stratification for $z > l$, resulting in a steep drop in the density.
This density drop is expected to support an f mode.

Figure \ref{fig: komega NF} shows $k$-$\omega$ diagrams at two heights, averaged over four realizations of the NF case.
We observe both p and f modes near the top of the domain, while g modes are also visible in the stably stratified overshoot region.
The expected dispersion relation of the f mode ($\omega = \sqrt{g k_x}$) has been indicated as a dashed grey line.
The deviation of the f mode frequencies from the theoretical curve at high $\sclk_x$ is due to the upper boundary of the domain ($z = 1.05 \, l$) being close to the density jump induced by the cooling layer ($z=l$).

\section{Strengthening of the f mode}
\label{section: f mode str}

\begin{figure}
	\centering
	\includegraphics{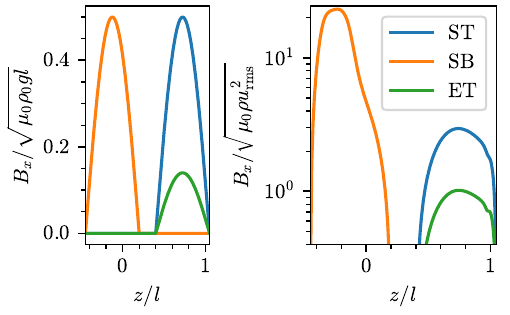}
	\caption{
	Profiles of $B_x$ in the ST and SB simulations.
	Recall that the chosen magnetic field is only a function of $z$.
	The profiles of $\rho$ and $u_\text{rms}$ used to normalize the magnetic field correspond to horizontal-temporal averages, just like the quantities shown in figure \ref{helio.conv-mag.hydro: fig: background + fluxes}.
	}
	\label{helio.conv-mag.SBT: fig: bx ST}
\end{figure}

\begin{figure}
	\centering
	\includegraphics{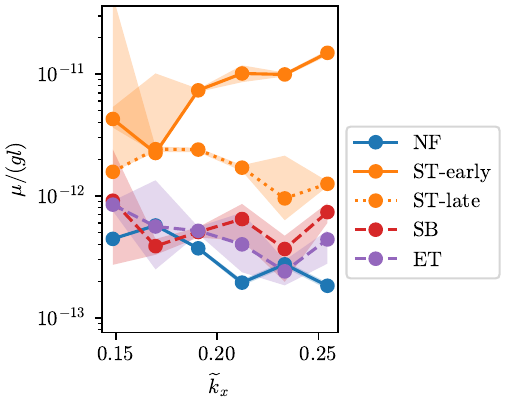}%
	\caption{
	The mass of the f mode at $z = l$ as a function of $\sclk_x$ with $\sclk_y = 0$.
	The two cases of the `ST' configuration are marked `early' and `late' to indicate the timespan considered (see table \ref{table: params and diagnostics}).
	}
	\label{helio.conv-mag.SBT: fig: fmode vs k with err}
\end{figure}

\begin{figure}
	\centering
	\includegraphics[width=\linewidth, keepaspectratio]{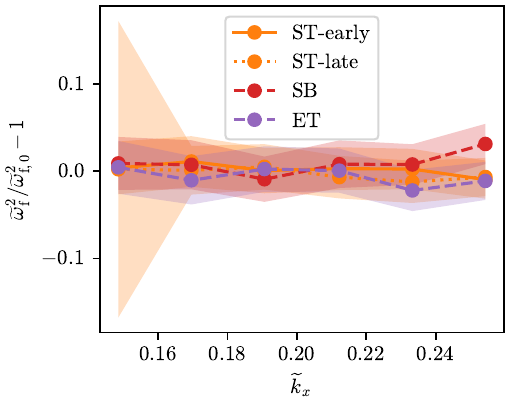}
	\caption{
	The shift in the frequency of the f mode (with respect to the NF case) at $z = l$ due to the imposed magnetic field, as a function of $\sclk_x$.
	The central frequency of the f mode in the magnetic case is $\sclom_\mathrm{f}$, while that in the non-magnetic case is $\sclom_{f,0}$
	}
	\label{helio.conv-mag.SBT: fig: fmode freq shift vs k}
\end{figure}

We now examine how simulations with imposed magnetic fields differ from the non-magnetic case described above.
Table \ref{table: params and diagnostics} shows the values of various parameters and diagnostics (the latter computed from a single realization) for each configuration.

The different magnetic field configurations are shown in figure \ref{helio.conv-mag.SBT: fig: bx ST}.
Recall that we have chosen magnetic fields such that $B_x$ is the only nonzero component, and it depends only on $z$.
The ST case is around 3.6 times the equipartition value, while the ET case is close to equipartition.
In both these cases, the magnetic field is nonzero near the top of the CZ.
On the other hand, the SB case has a magnetic field of the same strength as the ST case, but located below the CZ (and is thus highly super-equipartition).
The sensitivity of the f mode at the surface to magnetic fields at different depths can be gauged by comparing the ST and SB cases.
On the other hand, the ST and ET cases tell us about the effect of the strength of the magnetic field.
While these configurations are simpler than those expected for rising magnetic flux tubes, this study is a preliminary proof of concept.

Recall (section \ref{maghelio: section: mode model}) that we model the power spectrum at fixed $\sclk_x$, $\sclk_y$, and $z$ as the sum of a number of Lorentzians (for the modes) and a power law (for the continuum).
Since all the magnetic field configurations we impose do not have any spatial variation in the horizontal direction (unlike the fields one actually expects to find in the Sun), they lead to strong coupling between modes at $\sclk_x = \sclk_y = 0$ and those at nonzero wavenumbers (appendix \ref{appendix: mode coupling}).
In what follows, we confine ourselves to a range of $\sclk_x$ where the f mode does not seem to be affected by this effect.

Figure \ref{helio.conv-mag.SBT: fig: fmode vs k with err} shows the mode mass (defined in section \ref{section: mode mass defn}) corresponding to the f mode at $z = l$ as a function of $\sclk_x$ (with $\sclk_y = 0$) for all the magnetic field configurations we consider (along with the non-magnetic case).
The mode masses in the SB, ET, and NF cases are not significantly different.
On the other hand, the mode masses in the ST case are up to an order of magnitude higher.
However, the strengthening of the f mode in the ST case is transient: the mode masses measured immediately after the imposition of the magnetic field (ST-early) are significantly larger than those measured some time later (ST-late).

Figure \ref{helio.conv-mag.SBT: fig: fmode freq shift vs k} shows that there is no significant shift in the frequency of the f mode due to imposed magnetic fields.
The finiteness of the integration time (which constrains the minimum frequency difference that one can resolve) is the major source of error in the f mode frequencies,
and has thus been added in quadrature to the fitting errors for this figure alone.
While the lack of a significant shift is at variance with previous forced-turbulence studies that imposed a uniform magnetic field \citep[fig.~9]{SinBraChi15}, it is consistent with studies in which a non-uniform and localized magnetic field was imposed \citep{SinRaiKap20}.

We note that the point where the magnetic field peaks in the SB case is less than 6 pressure scale heights below the point where the f mode is measured.
Further, this depth is smaller than the depth over which the eigenfunction of the f mode is expected to decay exponentially (e.g.\@ the eigenfunction of the f mode for $\sclk_x = 0.25$ is expected to decay exponentially over a length scale of approximately $1.36 l$).

\section{Conclusions}
\label{section: conclusions}

In agreement with previous forced-turbulence studies \citep{SinRaiKap20}, we have found that a super-equipartition magnetic field localized near the surface causes strengthening of the f mode.
On the other hand, an otherwise identical magnetic field centered 6 pressure scale heights below the point of measurement does not seem to affect the amplitude of the f mode.
We also find no significant effect for an equipartition magnetic field localized near the surface.
None of the magnetic cases shows a significant shift in the frequency of the f mode as compared to the non-magnetic case.

Even in cases where the magnetic field strengthens the f mode, we have found that the strengthening of the f mode is transient, with the mode mass decreasing with time.
This means future studies that try to quantitatively constrain solar magnetic fields based on observed mode amplitudes must model both the time evolution and the spatial dependence of these fields.

Quantitative interpretation of observations \parencite{SinRaiBra16, WaiRotSin23} requires a better understanding of how such effects depend on the strength, the spatial scale, and the growth rate of the magnetic field.
A more extensive parameter study to this end is in progress.
Future work will also examine the effects of magnetic fields on p and g modes in simulations such as those presented here.
Similar simulations with rotation may aid the interpretation of observational reports of the cycle-dependence of the power and frequency of solar Rossby modes \citep{WaiZha23}.

\begin{acknowledgments}
	We thank Matthias Rheinhardt for fixing a bug in \pencil{} that affected our study.
	The Gauss Centre for Supercomputing e.V. (\url{www.gauss-centre.eu}) supported this project by providing computing time on the GCS Supercomputer SuperMUC-NG at Leibniz Supercomputing Centre (\url{www.lrz.de}).
	We also used the Pegasus HPC facility at IUCAA.
	We thank Alexandra Elbakyan for facilitating access to scientific literature.
\end{acknowledgments}

\vspace{5mm}

\textit{\large Data Availability:}
The input files, postprocessing scripts, and a subset of the output (enough to reproduce our results) for all the simulations described in this paper can be downloaded from Zenodo \parencite{zenodo_f_mag_str_2023}.
The available output includes horizontally averaged diagnostics and power spectra (the latter at $z=l$ for the magnetic cases, and $z = -0.1l,l$ for the NF cases).
Power spectra at other values of $z$ will be provided on reasonable request.

\software{
	\pencil{} \citep{Pencil2021},
	SciPy \citep{scipy2020},
	NumPy \citep{numpy2020},
	Matplotlib \citep{matplotlib2007},
	h5py \citep{h5py3.8.0},
	GNU Parallel \citep{gnuparallel2022}.
}

\appendix
\section{Grid independence}
\label{appendix: grid-independence}

\begin{figure*}
	\centering
	\includegraphics{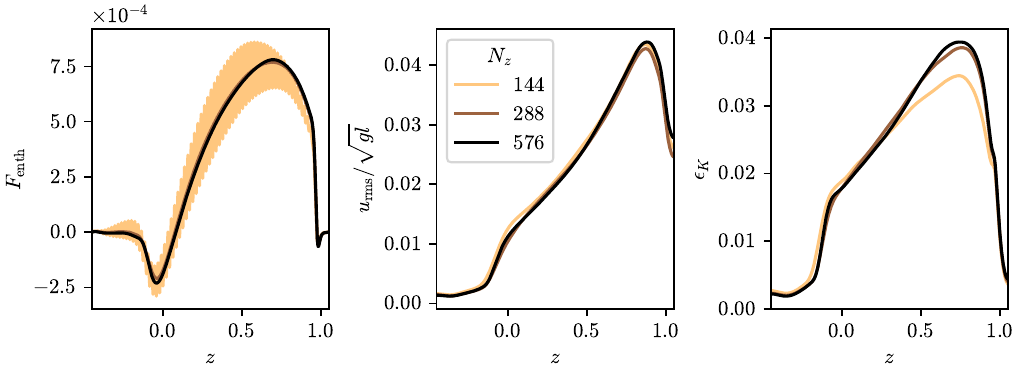}
	\caption{
		Dependence of various horizontally and temporally averaged quantities on the number of grid points.
		$N_x$ and $N_y$ are scaled by the same factor as $N_z$, with only the latter being indicated in the legend.
		$\epsilon_K \defn \meanBrxyt{ 2 \rho\nu S_{ij} S_{ij} }$ is the dissipation rate of the kinetic energy.
		Note that in the $N_z=144$ case, $F_\enth$ oscillates on the grid scale even after horizontal-temporal averaging.
		}
	\label{fig: grid independence convection}
\end{figure*}
\begin{figure}
	\centering
	\includegraphics{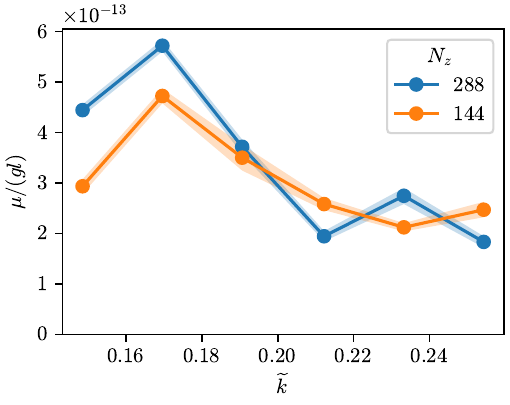}
	\caption{
		Dependence of the mass of the f mode on the grid resolution.
		$N_x$ and $N_y$ are scaled by the same factor as $N_z$, with only the latter being indicated in the legend.
		}
	\label{fig: grid independence mode mass}
\end{figure}

To check if the number of grid points is sufficient to resolve all the relevant scales, we consider a set of simulations with $L_x = L_y = 2l$ (all the other parameters are kept identical to those in the NF case).
We initialize a simulation with this aspect ratio (keeping the grid resolution unchanged) by tiling a snapshot from the $L_x = L_y = l$ simulation (described in section \ref{section: initial conditions}) and running it for $300 \, \sqrt{l/g}$ time units.
The resulting snapshot is then linearly interpolated onto grids of different resolutions, keeping the domain size unchanged.
These simulations are then run for $500 \, \sqrt{l/g}$ time units.
Various horizontal-temporal averages, calculated by averaging over the last $300 \, \sqrt{l/g}$ time units, are shown in figure \ref{fig: grid independence convection}.
The depicted quantities are converged to within a few percent for $N_z = 288$ (used for the simulations reported in this study).

We note that the mode masses we report are all measured at wavenumbers corresponding to length scales somewhat larger than the integral scale of the convective flows; for reference, a mode with $k = 2\pi/l$ would correspond to $\sclk \approx 0.34$.
We thus expect the mode masses to not be more sensitive to the resolution than the quantities shown in figure \ref{fig: grid independence convection}.
Figure \ref{fig: grid independence mode mass} shows that when the grid spacing is doubled compared to what we have used for our study, the mass of the f mode in the non-magnetic case does not change to an extent that would invalidate our conclusions.

\section{Mode coupling}
\label{appendix: mode coupling}

\begin{figure}
	\centering
	\includegraphics{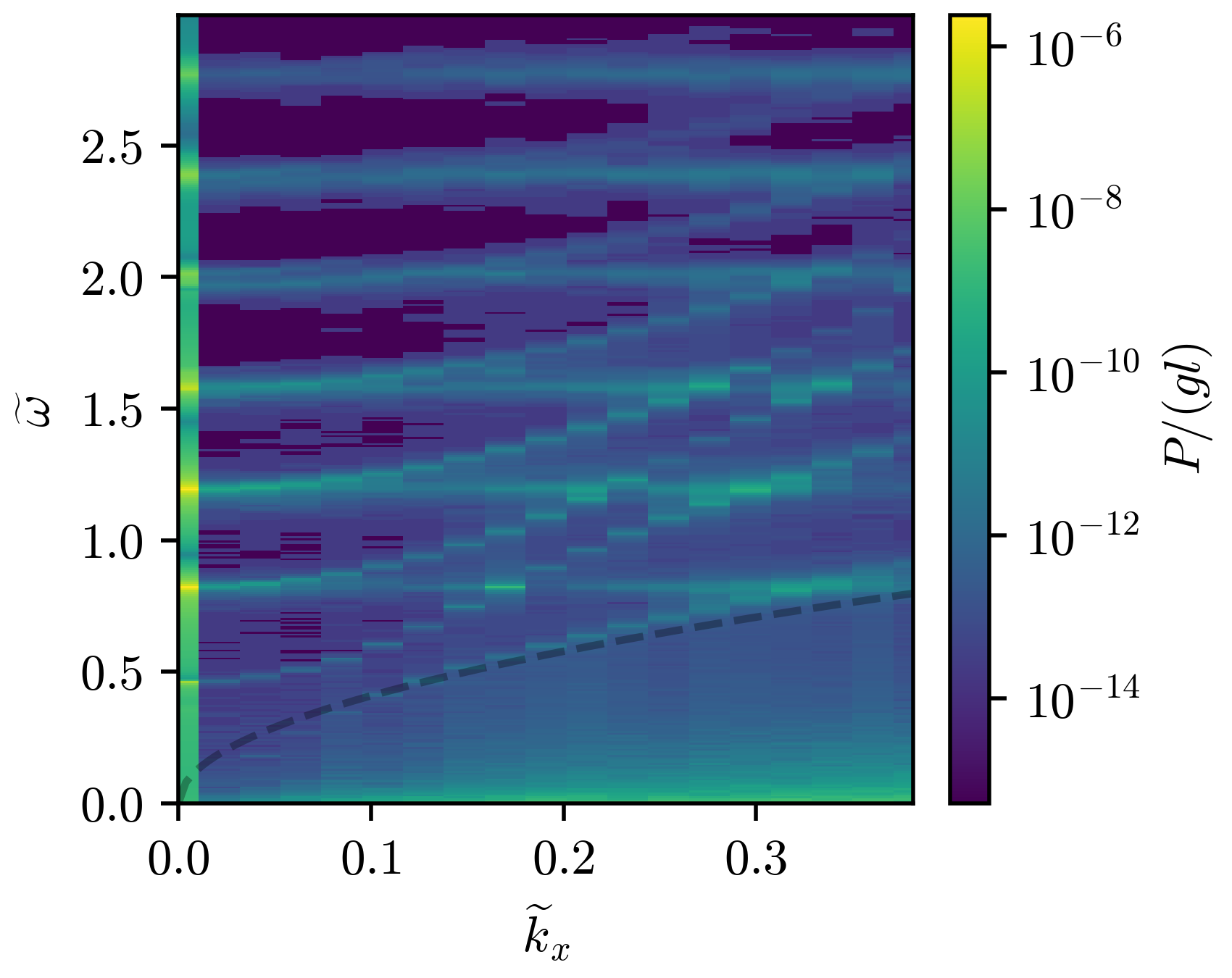}
	\caption{
		Diagnostic $k$-$\omega$ diagram generated by averaging over  all realizations of the ST-early case for $z=l$.
		In both cases, we have set $\sclk_y = 0$.
		The dashed black line shows the theoretical dispersion relation of the f mode.
		}
	\label{fig: komega ST}
\end{figure}

Figure \ref{fig: komega ST} shows a $k$-$\omega$ diagram from the ST-early case.
Unlike in the non-magnetic case (figure \ref{fig: komega NF}), one sees a vertical line at $\sclk_x = 0$.
Similar vertical features (referred to as \emph{Bloch modes}) have been observed in earlier work, at multiples of twice the wavenumber of the horizontal variation of the magnetic field (\citealp[p.~4]{SinBraRhe14}; \citealp[sec.~3.2]{SinRaiKap20}).
Recall that the magnetic field we have imposed has no spatial variation in the horizontal direction (corresponding to $\sclk_x = \sclk_y = 0$).
In simulations where the magnetic field varies sinusoidally in the horizontal direction (not reported here), we also observe such features at twice the corresponding wavenumber.
\Textcite[section 3.3]{SinRaiKap20} found that the Bloch modes are absent when the magnetic field is horizontally localized.
Since the magnetic fields in sunspots are localized, we do not believe the Bloch modes observed in our setup have any direct observational relevance.

Another difference from the non-magnetic case is the appearance of horizontal bands.
These appear to be caused by nonlinear interactions due to the large amplitudes of the p modes at $\sclk_x = 0$.
In line with our comments on Bloch modes, we ignore wavenumber ranges where the f mode is affected by this mode coupling.

\section{Fitting modes}
\label{appendix: fitting modes}

\begin{table*}
	\centering
	\begin{tabular}{
		D{.}{.}{1,2}
		D{.}{.}{1,2}
		D{.}{.}{1,2}
		D{.}{.}{1,2}
		c
		l
		c
		l
		}
		\toprule
		   \mcL{$\sclk_x$} &   \mcL{$\sclom_\text{{min}}$} &   \mcL{$\sclom_\text{{max}}$} &   \mcL{$\sclom_\text{{guess}}$} &   \mcL{$n_\text{{Lor}}$} & \mcL{$\sclom_\text{{extra}}$}   &   \mcL{$n_\text{{Lor,mag}}$} & \mcL{$\sclom_\text{{extra,mag}}$}   \\
		\midrule
			0.15 &                     0.2 &                    0.65 &                      0.52 &                  1 &                           &                      1 & 0.45                          \\
			0.17 &                     0.3 &                    0.7  &                      0.55 &                  1 &                           &                      1 & 0.45                          \\
			0.19 &                     0.3 &                    1    &                      0.6  &                  2 & 0.9                       &                      2 & 0.45; 0.82                    \\
			0.21 &                     0.3 &                    1.3  &                      0.63 &                  3 & 0.96; 1.15                &                      3 & 0.45; 0.82; 1.21              \\
			0.23 &                     0.3 &                    1.13 &                      0.67 &                  2 & 1.04                      &                      1 & 0.81                          \\
			0.25 &                     0.3 &                    1.4  &                      0.72 &                  3 & 1.08; 1.33                &                      2 & 0.82; 1.21                    \\
		\bottomrule
	\end{tabular}
	\caption{
		Parameters used to fit the f mode.
		Note that all the frequencies listed above are inputs for the fitting routine; these are not the same as the central frequencies of the modes in the final fit.
		}
	\label{helio.conv-mag.SBT: table: fmode guesses}
\end{table*}

The continuum is modelled as a power law,
\begin{equation}
	C_1 \abs{\sclom}^{-C_2}
	\,,
\end{equation}
with $C_1$ and $C_2$ constrained to be positive.
This is motivated by the fact that on dimensional grounds, the frequency spectrum of the turbulent velocity field is expected to scale as $E(\omega) \scalesAs \epsilon \omega^{-2}$, where $\epsilon$ is the dissipation rate of the specific kinetic energy.

Each mode is modelled as a Lorentzian,
\begin{equation}
	\frac{C_3 \gamma / \pi }{ \left( \sclom - \sclom_0 \right)^2 + \gamma^2 }
	\,,
\end{equation}
with $C_3$ constrained to be positive.
The half-width at half-maximum (HWHM, $\gamma$) is constrained to obey $0 < \gamma < \gamma_\text{max}$ with $\gamma_\text{max} = 0.05$.

For each value of $\sclk_x$ (recall that we set $\sclk_y = 0$ and $z = l$), $P(\sclom)$ in the interval $\sclom_\text{min} < \sclom < \sclom_\text{max}$ is considered (see table \ref{helio.conv-mag.SBT: table: fmode guesses}).
The bounds on $\sclom$ and the number of modes to be fit are determined by eye and kept fixed.
For all the Lorentzians, we require $\sclom_\text{min} + \gamma_\text{max} < \sclom_0 < \sclom_\text{max} - \gamma_\text{max}$.

For the non-magnetic case, we consider $n_\text{Lor}$ Lorentzians with the initial guesses for their frequencies being given by $\sclom_\text{guess}$ and $\sclom_\text{extra}$.
Due to the effects discussed in appendix \ref{appendix: mode coupling}, the magnetic cases have extra peaks of $P(\sclom)$ as compared to the non-magnetic cases.
$n_\text{Lor}$ and $\sclom_\text{extra}$ are thus augmented in the magnetic cases by $n_\text{Lor,mag}$ and $\sclom_\text{extra,mag}$ respectively.
Table \ref{helio.conv-mag.SBT: table: fmode guesses} gives the values of these parameters.

To reduce sensitivity to the initial parameter guesses, each mode is fit in two stages.
For the first stage, the initial guesses for the mode frequencies are set by eye, and $P(\sclom)$ is convolved along the frequency axis with a Lorentzian of HWHM $0.01$.
The considered range of $\sclom$ is widened by the same amount.
The resulting optimal parameters are then used as initial guesses to fit the unsmoothed $P(\sclom)$ in the second stage.

For the second stage, the errors in $P(\sclom)$ (estimated as described in section \ref{section: error estimation}) are smoothed along the frequency axis with a top hat of half-width $0.1$.
During the process of propagating these errors to estimate the errors in the fit parameters, a scaling factor is applied to the errors in the fit parameters such that the reduced $\chi^2$ of the fit to $P(\omega)$ becomes 1.

Among all the Lorentzians in the final fit, the f mode is identified as the Lorentzian with the largest mass among all the Lorentzians whose central frequency is within $0.05$ of $\sclom_\text{guess}$.

\section{On modelling the continuum}
\label{section: continuum power law slope}

As discussed in appendix \ref{appendix: fitting modes}, we have modelled the continuum (which is expected to be dominated by the spectrum of the turbulent flows themselves) as a power law.
The power law slope obtained from our simulations is consistent with the theoretically expected value of 2 (for the different ranges of $\sclom$ given in table \ref{helio.conv-mag.SBT: table: fmode guesses}, we find $1.75 \lesssim C_2 \lesssim 2$ for the non-magnetic cases, where $C_2$ is the negative power law index; the magnetic cases show a larger spread in the power law indices, likely due to stochastic errors associated with the much smaller time intervals considered).

There seems to be no consensus on what functional form the continuum should take.
In previous studies, the continuum was modelled as a polynomial \parencite{SinRaiBra16}; as a constant \parencite{SinBraChi15, SinRaiKap20}; or as the sum of a constant and a Gaussian \parencite{WaiRotSin23}.
\Textcite{KorKorOls22} stretched the assumption of a constant background even further, by treating the average of the power spectrum at low frequencies as the continuum at the frequencies of interest.

Recall that \textcite{SinRaiBra16} and \textcite{WaiRotSin23} found that the f mode is strengthened prior to the emergence of active regions.
On the other hand, \textcite{KorKorOls22}, found no significant strengthening.
One possible explanation for this disagreement is the differing treatment of the continuum.
Nevertheless, since their analyses differ in other respects too, further study is needed to resolve this issue.

\bibliography{refs.bib}
\bibliographystyle{aasjournal}

\end{document}